\documentclass[]{ceurart}

\usepackage{minted}
\setminted{breaklines=true}

\usepackage{hyperref}
\usepackage{cleveref}

\begin{document}

\copyrightyear{2026}
\copyrightclause{Copyright for this paper by its authors.
  Use permitted under Creative Commons License Attribution 4.0
  International (CC BY 4.0).}

\conference{Joint Proceedings of REFSQ-2026 Workshops, Doctoral Symposium, Posters \& Tools Track, and Education and Training Track. Co-located with REFSQ 2026. Poznan, Poland, March 23-26, 2026}

\title{Towards a Software Reference Architecture for Natural Language Processing Tools in Requirements Engineering}

\author[1]{Julian Frattini}[
    email=julian.frattini@chalmers.se, 
    orcid=0000-0003-3995-6125, 
    url=https://julianfrattini.github.io/,
]
\address[1]{Chalmers University of Technology and University of Gothenburg, Sweden}

\author[2]{Quim Motger}[%
    orcid=0000-0002-4896-7515,
    email=joaquim.motger@upc.edu,
    url=https://quim-motger.github.io/
]
\address[2]{Universitat Politècnica de Catalunya, Spain}

\begin{abstract}
    % Context
    Natural Language Processing (NLP) tools support requirements engineering (RE) tasks like requirements elicitation, classification, and validation. 
    % Gap
    However, they are often developed from scratch despite functional overlaps, and abandoned after publication.
    This lack of interoperability and maintenance incurs unnecessary development effort, impedes tool comparison and benchmarking, complicates documentation, and diminishes the long-term sustainability of NLP4RE tools.
    % Goal
    To address these issues, we postulate a vision to transition from monolithic NLP4RE tools to an ecosystem of reusable, interoperable modules.
    % Method
    We outline a research roadmap towards a software reference architecture (SRA) to realize this vision, elaborated following a standard methodological framework for SRA development. 
    As an initial step, we conducted a stakeholder-driven focus group session to elicit generic system requirements for NLP4RE tools.
    % Results
    This activity resulted in 36 key system requirements, further motivating the need for a dedicated SRA.
    % Conclusions
    Overall, the proposed vision, roadmap, and initial contribution pave the way towards improved development, reuse, and long-term maintenance of NLP4RE tools.
\end{abstract}


\maketitle

\section{Introduction}
\label{sec:intro}

The prevalence of natural language (NL) in requirements specifications makes natural language processing (NLP) techniques attractive to automate requirements engineering (RE) tasks~\cite{Zhao2022}. 
This gave rise to the development of NLP4RE tools, which are ``\emph{any software (e.g., a script, executable, or web service) that employs NLP technology to support one or more RE activities}''~\cite{Frattini2025}.
Examples include tools for requirements extraction~\cite{Motger2025}, traceability link recovery~\cite{hey2024requirements}, and test case generation~\cite{fischbach2023automatic}.

Despite their performance, academic NLP4RE tools exhibit several problems, including a lack of longevity~\cite{frattini2024requirements}, superficial documentation~\cite{Abualhaija2024}, and low reusability in industrial and academic contexts~\cite{Zhao2022}.
Many of these problems are connected to the absence of architectural guidance~\cite{frattini2024requirements}, which forces developers of NLP4RE tools to repeatedly re-implement common software steps such as input document parsing, model integration, result visualization, and data formatting for evaluation or benchmarking. 
%As a result, tools are designed ad-hoc with incompatible data representations, limiting reuse, comparison, and integration across the NLP4RE ecosystem. 

In this paper, we summarize the challenges noted in prior systematic studies (\Cref{sec:background}) and formulate our vision to overcome them (\Cref{sec:vision}).
In the roadmap to achieve this vision (\Cref{sec:roadmap}) we pick up the suggestion from prior work that a shared software reference architecture (i.e., a high-level blueprint defining common structures, components, and principles to guide system design~\cite{Garces2021}) can address these challenges~\cite{Frattini2025}. 
As the first step of this roadmap, we report results from a focus group activity with NLP4RE stakeholders (\Cref{sec:preliminary}), aimed at (1) paving the way towards this vision, (2) supporting its feasibility, and (3) guiding the elicitation of architectural requirements. 
All materials derived from this study are publicly available at \url{https://github.com/airera/study-sra} and archived at \url{https://doi.org/10.5281/zenodo.18268705}.

\section{Background}
\label{sec:background}

NLP4RE tools enjoy great popularity~\cite{Zhao2022} yet lack established guidance for their design, development, and maintenance~\cite{Frattini2025}. 
While the community is adopting standard frameworks for areas such as documentation~\cite{Abualhaija2024}, far less attention has been given to common, canonical guidelines for design and development.
Though diverse in their actual implementation, NLP4RE tools share several functional and qualitative aspects. 
For instance, they share multiple steps of their processing pipeline, such as: reading an input document, pre-processing textual requirement artifacts, analyzing them, and producing some form of output through a user interface~\cite{Zhao2022}.
On the other hand, these NLP4RE tools often support similar RE activities (e.g., requirements elicitation or analysis) using similar task types (e.g., classification or tracing \& relating)~\cite{Frattini2025}. 
Furthermore, qualitative aspects such as explainability of results, interoperability of software components and recoverability of experimental results are recurrent concerns across NLP4RE tools~\cite{beqiri2024classifying,arrabito2020comparison}.
Still, NLP4RE tools are developed in isolation as monoliths, inducing several challenges:
(1) Despite functional overlaps, NLP4RE tools \textbf{rarely reuse} any existing implementations, forcing the developer to build all parts of the tool, not just the uniquely new parts.
(2) NLP4RE tools are mostly academic software, which is often \textbf{constrained to very specific use cases} (e.g., only able to read one type of input document).
(3) Given the diverse architectures of tools with similar use cases, it is \textbf{difficult to compare} them against a common benchmark.
(4) Most NLP4RE tools are \textbf{not maintained} after publication, making them prone to break soon after their development.

%\begin{itemize}
%    \item \textbf{Limited reuse}: Despite functional overlaps, NLP4RE tools hardly reuse any existing implementations, forcing the developer to build all parts of the tool, not just the uniquely new parts.
%    \item \textbf{Limited applicability}: NLP4RE tools are mostly academic software, which is often constrained to very specific use cases (e.g., only able to read one type of input document).
%    \item \textbf{Limited comparability}: Given the diverse architectures of tools with similar use cases, it is difficult to compare them against a common benchmark.
%    \item \textbf{Limited maintenance}: Most NLP4RE tools are discarded after publication, making them prone to break soon after their development.
%\end{itemize}

\section{Vision}
\label{sec:vision}

We envision that the landscape of NLP4RE tools would benefit from a shift away from isolated, monolithic tools towards \emph{an ecosystem of interoperable and reusable modules}. 
This ecosystem would address the aforementioned issues as follows:

\begin{enumerate}
    \item \textbf{Improved reuse}: If interoperable modules encapsulating functionality of common steps of the pipeline of an NLP4RE tool were available, new tools could simply reuse appropriate, existing modules. For example, several NLP4RE tools start by ingesting, formatting, and cleaning a requirements artifact in a \emph{csv} format. If this functionality were encapsulated in an interoperable \emph{input parser} module, all these tools could reuse that same module.
    This would reduce the development effort to only the new modules and their composition with the reused ones. 
    \item \textbf{Improved applicability}: Interoperable modules would also extend the applicability of NLP4RE tools beyond specific use cases and fixed input-output combinations. For example, an input parser accepting only \emph{csv} files could be replaced with another module accepting \emph{docx} files without modifying the rest of the tool if both parsed the input document to a common internal representation of the requirements. 
    \item \textbf{Improved comparability}: If NLP4RE tools followed a common modular structure, tools addressing the same RE activity and with the same task type could be compared easily on dedicated benchmarks. For example, alternative classification modules could be evaluated in isolation while keeping the rest of the pipeline unchanged.
    \item \textbf{Improved maintenance}: Shifting from isolated, standalone tools to a common ecosystem of modules would distribute maintenance responsibilities across contributing authors. For example, widely reused modules could benefit from joint maintenance efforts, increasing their longevity.
\end{enumerate}

\begin{figure}[ht]
    \centering
    \includegraphics[width=\linewidth]{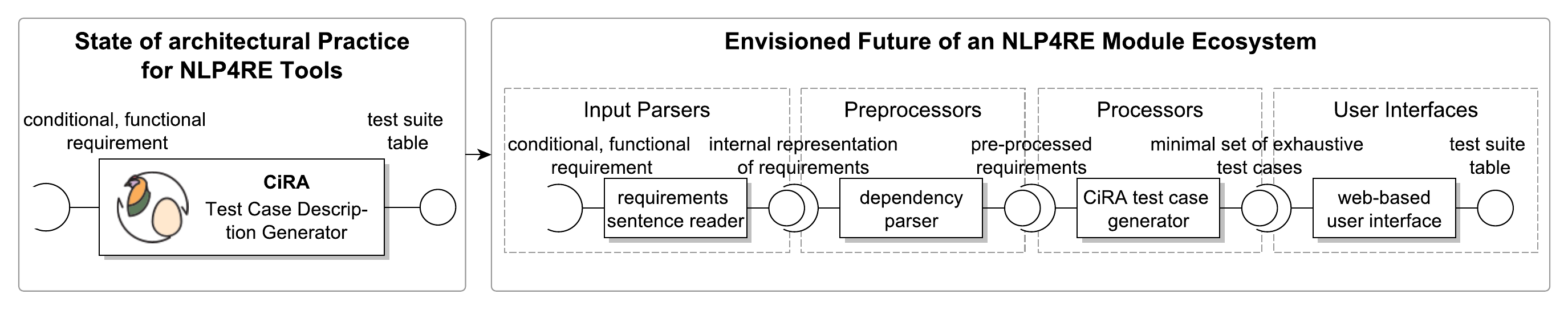}
    %\vspace{-.6cm}
    \caption{State of practice and envisioned future of NLP4RE tool architectures}
    \label{fig:vision}
\end{figure}

\Cref{fig:vision} illustrates this paradigm shift. 
On the left, tools such as \emph{CiRA}~\cite{frattini2023cira} are shown as monolithic systems with highly coupled \emph{input}, \emph{processing logic}, and \emph{output} functionality within a single tool boundary. 
On the right, these are decomposed into interoperable modules, including \emph{Input Parsers} (e.g., \emph{requirements sentence reader} parsing requirements into a common internal representation), \emph{Pre-processors}, \emph{Processors} (e.g., the \emph{CiRA test case generator}), and \emph{User Interfaces}. 
This separation enables individual modules to be independently reused, evaluated, maintained, replaced, and composed across NLP4RE tools, while preserving flexibility in tool configuration and evolution.

\section{Roadmap}
\label{sec:roadmap}

To achieve the vision outlined in \Cref{sec:vision}, we propose the elaboration of a software reference architecture (SRA) as a means to guide the development of interoperable, reusable, and consistent NLP4RE tools.
An SRA is an abstraction of domain-specific software architectures guiding and standardizing system design and development practices within that domain~\cite{Garces2021}. 
Related work started to address similar problems by building architectural standards and reference models within the NLP4RE community. 
As an example, D\k{a}browski et al. designed a reference model and architecture for review-based user feedback mining within the context of software engineering activities~\cite{dkabrowski2022mining}. 
More recently, D\k{a}browski et al. discussed early results on the analysis towards an SRA for agentic RE systems~\cite{Dabrowski2025IntelligentAgents}. Despite this, and to the best of our knowledge, no SRAs have been proposed within the context of NLP4RE tools.

\begin{figure}[ht]
    \centering
    \includegraphics[width=\linewidth]{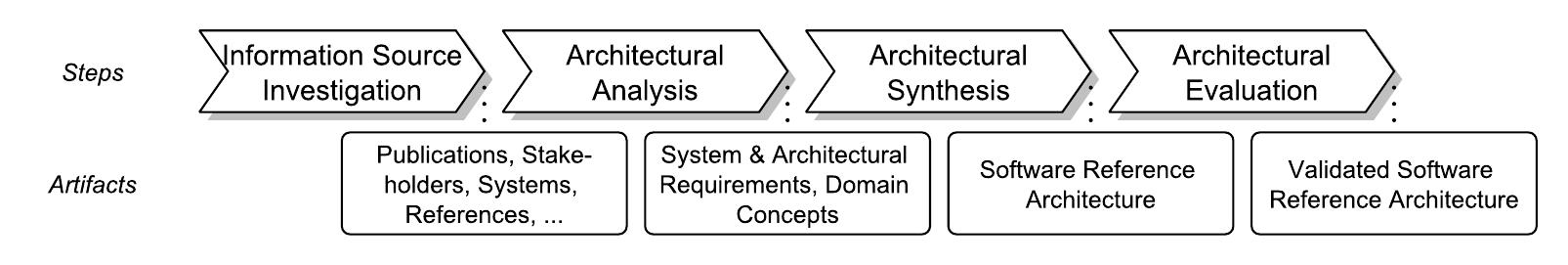}
    %\vspace{-.6cm}
    \caption{Four step ProSA-RA method for developing SRAs by Nakagawa et al.~\cite{Nakagawa2014}}
    \label{fig:roadmap}
\end{figure}

To elaborate such SRA, we propose a roadmap following the synthesized guidelines of Nakagawa et al. for designing and evaluating an SRA~\cite{Nakagawa2014}.
The guidelines consist of four steps visualized in \Cref{fig:roadmap} and elaborated in \Cref{sec:roadmap:investigation,sec:roadmap:analysis,sec:roadmap:synthesis,sec:roadmap:evaluation}.

\subsection{Information Source Investigation}
\label{sec:roadmap:investigation}

Nakagawa et al.~\cite{Nakagawa2014} propose to initiate the SRA design by gathering information sources to elicit requirements from.
The five types of information map to the domain of NLP4RE tools as follows:

\begin{itemize}
    \item \textbf{Publications.} Systematic mapping studies on articles applying NLP to RE activities, such as information extraction, classification, and traceability~\cite{Zhao2022,Umar2024,Necula2024}.
    \item \textbf{Software systems (tools).} Descriptions, classifications and characterization of existing NLP4RE tools that provide insights into architectural decisions and integration practices~\cite{Frattini2025}.
    \item \textbf{Reference models and architectures.} Prior reference models and architectural frameworks that serve as conceptual foundations for designing domain-specific NLP4RE SRAs~\cite{dkabrowski2022mining,Dabrowski2025IntelligentAgents}.
    \item \textbf{Ontologies.} Domain and requirements ontologies that formalize core concepts, e.g., the OpenReq ontology~\cite{OpenReqOntology} and NLP4RE-focused taxonomies for documentation of tools~\cite{Abualhaija2024}.
    \item \textbf{Stakeholders.} Experts and practitioners involved in the design, development, or use of NLP4RE tools who contribute their experiences and perspectives. 
\end{itemize}

\subsection{Architectural Analysis}
\label{sec:roadmap:analysis}

From the identified information sources, three types of requirements are to be elicited~\cite{Nakagawa2014}:

\begin{enumerate}
    \item \textbf{System requirements (SR)}: Key functional and non-functional requirements of existing and envisioned NLP4RE systems. 
    \item \textbf{Architectural requirements (AR)}: A selection and abstraction of the SRs which can be delegated to an SRA. Multiple SRs may map to a single AR when several system-level needs converge into one architectural concern. The resulting set of ARs represents requirements that an SRA must meet such that any NLP4RE tool designed based on the SRA exhibits the selected SRs.
    \item \textbf{Domain concepts (DC)}: Concepts capturing the main architectural entities and relationships to build a reference model of the NLP4RE domain (containing, for example, data sources, processing components, reasoning layers).
\end{enumerate}

\subsection{Architectural Synthesis}
\label{sec:roadmap:synthesis}

Next, Nakagawa et al. suggest to implement an SRA that fulfills the elicited ARs~\cite{Nakagawa2014}.
%We adjust this step with one intermediate consideration: After the elicitation of ARs but before the development of an according SRA, we suggest to assess whether there an SRA that would meet all ARs already exists. Only if no SRA exists that meets all ARs is the development of a new SRA warranted. Should no existing SRA meet all elicited ARs, a
An SRA shall be specified in terms of architectural views~\cite{clements2011documenting}, i.e., models of the software architecture from several different perspectives.
Nakagawa et al. recommend specifying the architecture from the viewpoint of the design-time modules, runtime components, and deployment on hardware~\cite{Nakagawa2014}.
These views shall be described in diagrams (e.g., UML) and textual explanations to guide developers in the design of their NLP4RE tool.
For example, interface specifications would define a common output for all implementations of \textit{input parser} modules to ensure their exchangeability as the initial module of a tool.
An open platform to host, share, reuse, and evolve contributions can facilitate the adoption and use of the SRA.
Distilling guidelines and structured advice in the form of tutorials or handbooks~\cite{Frattini2025} would further improve the adoption of the SRA.

\subsection{Architectural Evaluation}
\label{sec:roadmap:evaluation}

Once developed, the completeness, usefulness, and adoption potential of the SRA feedback needs to be evaluated through expert feedback, tool-based experimentation, and community engagement.
For the NLP4RE domain, we envision the following evaluations:

\begin{itemize}
    \item \textbf{Traceability Matrix Mapping}: As done by D\k{a}browski et al.~\cite{Dabrowski2022}, mapping architectural modules and components as specified in the SRA to the elements of existing NLP4RE tools shows its general applicability and validates its usefulness by highlighting gaps.
    \item \textbf{Perceived Usefulness}: Surveying NLP4RE tool authors regarding the perceived usefulness, ease of use, and intention to use the SRA validates its applicability.
    \item \textbf{Tool Refactoring}: Refactoring previously published tools according to the SRA will demonstrate its applicability and contribute initial reusable, interoperable modules. First, internal refactoring of existing tools developed by the authors of the SRA~\cite{Motger2025,frattini2023cira} will verify that the SRA satisfies the elicited system requirements. Subsequently, external refactoring of third-party NLP4RE tools with community stakeholders will assess whether the SRA meets users’ goals in practice.

\end{itemize}

\section{Preliminary Results}
\label{sec:preliminary}

To further support our vision and as an initial seed towards architectural requirements elicitation, we contribute to the first step of our research roadmap (\Cref{sec:roadmap:investigation}) by eliciting information from stakeholders within the NLP4RE community, the only type of information source not readily available (other than publications, systems, etc.).
In this section, we summarize the outcomes of a stakeholder-driven focus group activity. 

\subsection{Method}
\label{sec:preliminary:method}

Aiming at interacting with researchers and practitioners familiarized with NLP4RE tools, we conducted a focus group session at the \emph{12th International Workshop on Artificial Intelligence and Requirements Engineering\footnote{https://aire-ws.github.io/aire25/}} (AIRE’25), attended by a sample of our target stakeholders. 
In a 45-minute activity, 20 workshop participants jointly elicited requirements across four categories: (i) functional requirements, (ii) non-functional requirements, (iii) domain concepts, and (iv) challenges related to the use of generative AI in NLP4RE systems, organized in two iterations.

In the first iteration, participants were divided into four groups, proportionally distributed across categories, each facilitated by a dedicated moderator. 
Each group was assigned one category and asked to discuss relevant requirements, concepts, or topics, which were captured as sticky notes on a flipboard. 
To foster prioritization and discussion, participants collaboratively positioned the notes in a two-dimensional matrix ranking \emph{priority} (low to high) and \emph{complexity} (low to high).

After the first iteration, participants from each group---except for the moderator---were redistributed across the remaining flipboards, forming new groups. 
Moderators then introduced and explained the existing items, after which participants could rearrange notes along the priority and complexity dimensions, and contribute new items. 
Finally, all participants reconvened, and moderators summarized the outcomes of each category, followed by a brief plenary discussion to clarify interpretations and exchange perspectives. 
The two authors of this study subsequently consolidated and thematically analyzed the collected information to synthesize and structure the results, serving as the basis for the elicitation of generic system requirements for NLP4RE tools.
Our replication package contains additional insights into the instructions provided to participants as well as generated results, including the raw data on the flipcharts and consolidated artifacts such as system-level requirements.

\subsection{Results and discussion}
\label{sec:preliminary:results}

The focus group activity resulted in a consolidated set of system-level requirements\footnote{The identifiers \emph{SRxy} refer to system-level requirements as listed in the replication package.} that directly reflect the four improvement dimensions outlined in our vision (Section~\ref{sec:vision}):

\begin{enumerate}
    \item \textbf{Improved reuse.} Participants emphasized externalizing common pipeline functionality to enable reuse across NLP4RE tools. This is reflected in requirements for modularity and component reuse (SR11), black-box interfaces (SR07), reusable input parsers for common formats (SR02, SR38), and transparent intermediate processing steps (SR06).

    \item \textbf{Improved applicability.} Stakeholders highlighted the need to overcome narrow, tool-specific use cases. Elicited requirements stress configurability and adaptability, including multilingual support (SR04, SR05), processing of specific RE artifacts (SR03), and explicit communication of input limitations (SR25).

    \item \textbf{Improved comparability.} Systematic comparison of NLP4RE tools emerged as a key concern. Participants called for evaluation support (SR01), reproducible behavior under fixed conditions (SR17), version tracking (SR18), and durable storage of artifacts and results (SR15), enabling controlled benchmarking of alternative processing modules.

    \item \textbf{Improved maintenance.} Long-term sustainability was addressed in requirements for modular design (SR10), version control (SR18), robustness and recoverability (SR14, SR15), and legal and data protection compliance (SR08, SR09), supporting shared maintenance of reusable modules.
\end{enumerate}

The elicited requirements provide stakeholder-grounded evidence that a modular, reference-architecture-driven approach can systematically address recurring challenges in NLP4RE tool development.

\section{Conclusions}
\label{sec:conclusions}

In this paper, we articulated a vision for transitioning NLP4RE tools from isolated, monolithic implementations towards an ecosystem of reusable and interoperable modules
We motivated this shift through a summary of challenges elicited in prior, systematic studies of the field~\cite{Zhao2022,frattini2024requirements,Frattini2025}.
We outlined a research roadmap grounded in established SRA development guidelines~\cite{Nakagawa2014} and provided early empirical support through a stakeholder-driven requirements elicitation activity, resulting in a consolidated set of system-level requirements for NLP4RE tools. 
Together, these contributions establish a concrete foundation for the design of a dedicated software reference architecture that can guide future tool development and evaluation. 
As future work, we will derive architectural requirements from the elicited system requirements, assess existing architectures against them, and iteratively design, implement, and evaluate an SRA in close collaboration with the NLP4RE community.

\begin{acknowledgments}
    We thank the participants of the focus group at the AIRE'25 workshop for their valuable insights.
  %This work was supported by TBD.
\end{acknowledgments}

\section*{Declaration on Generative AI}
\label{sec:generative-ai}

The authors employed generative AI tools, specifically GPT-5.2, for language editing and clarity improvements. 
All ideas, methods and interpretations remain the sole work of the authors.

\bibliography{references}

@article{Abualhaija2024,
  author = {Abualhaija, S. and others},
  title = {Replication in requirements engineering: The {NLP} for {RE} case},
  journal = {ACM Transactions on Software Engineering and Methodology},
  volume = {33},
  number = {6},
  pages = {Article 151, 33 pages},
  year = {2024},
  doi = {10.1145/3658669}
}

@article{Zhao2022,
  author = {Zhao, L. and others},
  title = {Natural language processing for requirements engineering: A systematic mapping study},
  journal = {ACM Computing Surveys},
  volume = {54},
  number = {3},
  pages = {Article 55, 1--41},
  year = {2022},
  doi = {10.1145/3444689}
}

@incollection{Frattini2025,
  author = {Frattini, J. and Unterkalmsteiner, M. and Fucci, D. and Mendez, D.},
  title = {{NLP4RE} Tools: Classification, Overview and Management},
  booktitle = {Handbook on Natural Language Processing for Requirements Engineering},
  editor = {Ferrari, A. and Ginde, G.},
  publisher = {Springer, Cham},
  year = {2025},
  doi = {10.1007/978-3-031-73143-3_13}
}

@article{Umar2024,
  author = {Umar, M. A. and Lano, K.},
  title = {Advances in automated support for requirements engineering: A systematic literature review},
  journal = {Requirements Engineering},
  volume = {29},
  number = {2},
  pages = {177--207},
  year = {2024},
  doi = {10.1007/s00766-023-00411-0}
}

@article{Necula2024,
  author = {Necula, S.-C. and Dumitriu, F. and Greavu-Șerban, V.},
  title = {A Systematic Literature Review on Using Natural Language Processing in Software Requirements Engineering},
  journal = {Electronics},
  volume = {13},
  number = {11},
  pages = {2055},
  year = {2024},
  doi = {10.3390/electronics13112055}
}

@article{Garces2021,
  author = {Garcés, L. and others},
  title = {Three decades of software reference architectures: A systematic mapping study},
  journal = {Journal of Systems and Software},
  volume = {179},
  pages = {111004},
  year = {2021},
  doi = {10.1016/j.jss.2021.111004}
}

@inproceedings{Nakagawa2014,
  author = {Nakagawa, E. Y. and others},
  title = {Consolidating a process for the design, representation, and evaluation of reference architectures},
  booktitle = {2014 IEEE/IFIP Conference on Software Architecture},
  pages = {143--152},
  year = {2014},
  doi = {10.1109/WICSA.2014.25}
}

@inproceedings{Dabrowski2025IntelligentAgents,
  author    = {Jacek D{\k{a}}browski and others},
  title     = {Intelligent Agents for Requirements Engineering: Use, Feasibility and Evaluation},
  booktitle = {IEEE 33rd International Requirements Engineering Conference (RE)},
  year      = {2025},
  doi       = {10.1109/RE63999.2025.00064}
}

@article{fischbach2023automatic,
  title={Automatic creation of acceptance tests by extracting conditionals from requirements: {NLP} approach and case study},
  author={Fischbach, Jannik and others},
  journal={Journal of Systems and Software},
  volume={197},
  pages={111549},
  year={2023},
  publisher={Elsevier},
  doi={10.1016/j.jss.2022.111549}
}

@inproceedings{hey2024requirements,
  title={Requirements classification for traceability link recovery},
  author={Hey, Tobias and Keim, Jan and Corallo, Sophie},
  booktitle={2024 IEEE 32nd International Requirements Engineering Conference (RE)},
  pages={155--167},
  year={2024},
  organization={IEEE},
  doi={10.1109/RE59067.2024.00024}
}

@article{frattini2024requirements,
  title={Requirements quality research artifacts: Recovery, analysis, and management guideline},
  author={Frattini, Julian and others},
  journal={Journal of Systems and Software},
  pages={112120},
  year={2024},
  publisher={Elsevier},
  doi={10.1016/j.jss.2024.112120}
}

@inproceedings{frattini2023cira,
  title={{CiRA}: An open-source Python package for automated generation of test case descriptions from natural language requirements},
  author={Frattini, Julian and Fischbach, Jannik and Bauer, Andreas},
  booktitle={2023 IEEE 31st International Requirements Engineering Conference Workshops (REW)},
  pages={68--71},
  year={2023},
  organization={IEEE},
  doi={10.1109/REW57809.2023.00019}
}

@inproceedings{dkabrowski2022mining,
  title={Mining user feedback for software engineering: Use cases and reference architecture},
  author={D{\k{a}}browski, Jacek and others},
  booktitle={IEEE 30th International Requirements Engineering Conference (RE)},
  pages={114--126},
  year={2022},
  doi={10.1109/RE54965.2022.00017}
}

@article{Motger2025,
  author    = {Quim Motger and others},
  title     = {Leveraging Encoder-Only Large Language Models for Mobile App Review Feature Extraction},
  journal   = {Empirical Software Engineering},
  year      = {2025},
  volume    = {30},
  number    = {4},
  pages     = {104},
  doi       = {10.1007/s10664-025-10660-y},
  issn      = {1573-7616}
}

@INPROCEEDINGS{Dabrowski2022,
  author={Dąbrowski, Jacek and others},
  booktitle={IEEE 30th International Requirements Engineering Conference (RE)}, 
  title={Mining User Feedback For Software Engineering: Use Cases and Reference Architecture}, 
  year={2022},
  volume={},
  number={},
  pages={114-126},
  doi={10.1109/RE54965.2022.00017}}

@techreport{OpenReqOntology,
  author       = {UH and UPC},
  title        = {OpenReq Ontologies and Patterns Catalogue},
  institution  = {OpenReq Project},
  year         = {2018},
  number       = {D5.3},
  address      = {European Commission, H2020 Programme},
  url          = {https://openreq.eu/},
  type         = {Deliverable},
}

@book{clements2011documenting,
	title={Documenting software architectures: views and beyond},
	author={Clements, Paul and Bachmann, Felix and Bass, Len and Garlan, David and Ivers, James and Little, Reed and Merson, Paulo and Nord, Robert and Stafford, Judith},
	year={2010},
	publisher={Addison Wesley},
	isbn={978-0321552686}
}

@inproceedings{beqiri2024classifying,
  title={Classifying ambiguous requirements: an explainable approach in railway industry},
  author={Beqiri, Lodiana and Montero, Calkin Suero and Cicchetti, Antonio and Kruglyak, Andrey},
  booktitle={2024 IEEE 32nd International Requirements Engineering Conference Workshops (REW)},
  pages={12--21},
  year={2024},
  organization={IEEE}
}

@inproceedings{arrabito2020comparison,
  title={A comparison of NLP Tools for RE to extract Variation Points.},
  author={Arrabito, Monica and Fantechi, Alessandro and Gnesi, Stefania and Semini, Laura and others},
  booktitle={REFSQ Workshops},
  year={2020}
}

@String{Computing = "Computing" }

@String{Springer = "Springer-Verlag" }

\end{document}